\documentclass[]{JHEP3} 

\usepackage{epsfig,multicol,bbm}

\providecommand{\tabularnewline}{\\}

\newcommand\science[3]   {\href{http://www.sciencemag.org/cgi/search?volume=#1&firstpage=#3}
		{{\it Science }{\bf #1} (#2) #3}}
\newcommand\etal {{\it et al.}}

\preprint{arXiv:1006:xxxx}

\title{Construction of a Kinematic Variable Sensitive to the Mass of the Standard Model Higgs Boson 
in $H\rightarrow WW^{*}\rightarrow\ell^{+}\nu\ell^{-}\bar{\nu}$
using Symbolic Regression}

\author{Suyong Choi\\
Department of Physics\\
Korea University, Seoul 136-713\\
Republic of Korea\\
E-mail: \email{suyong@korea.ac.kr}}

\received{\today} 		
\accepted{\today}		

\preprint{}

\abstract{
We derive a kinematic variable that is sensitive to the mass of the
Standard Model Higgs boson ($M_{H}$) in the $H\rightarrow WW^{*}\rightarrow\ell^{+}\nu\ell^{-}\bar{\nu}$
channel using symbolic regression method. Explicit mass reconstruction
is not possible in this channel due to the presence of two neutrinos
which escape detection. Mass determination problem is that of finding
a mass-sensitive function that depends on the measured observables.
We use symbolic regression, which is an analytical approach to the
problem of non-linear regression, to derive an analytic formula sensitive
to $M_{H}$ from the two lepton momenta and the missing transverse
momentum. Using the newly-derived mass-sensitive variable, we expect
Higgs mass resolutions between 1 to $4\ {\rm GeV}$ for $M_{H}$ between
130 and $190\ {\rm GeV}$ at the LHC with 10 $fb^{-1}$ of data. This
is the first time symbolic regression method has been applied to a
particle physics problem. 
}


\begin{document}

\section{Introduction}

In light of the current limits on the Standard Model (SM) Higgs boson mass ($M_H$), $H\rightarrow WW^{*}\rightarrow\ell^{+}\nu\ell^{-}\bar{\nu}$
channel is expected to be one of the most important channels in the search for the
Higgs boson \cite{LEPHiggs,CDFHWW,D0HWW,LEPewwg}.
Direct search results at the CERN LEP $e^{+}e^{-}$ collider places
a lower limit of 114.4 GeV on $M_H$ at 95\% confidence
level (C.L.) \cite{LEPHiggs}. And indirect constraints obtained from
fits to precision electroweak data, when combined with direct searches
at LEP, place an upper bound of 157 GeV at 95\% C.L. \cite{LEPewwg}.
In this mass range, the branching fraction of $H\rightarrow WW$ is
sizable and the production of $gg\rightarrow H$ through top quark loop
has the largest cross section for both the Tevatron and the LHC energies
\cite{SMHiggs}. Discovery of the Higgs boson and measurement of its properties 
are important for completing the picture of electroweak symmetry breaking
mechanism. However, measurement of mass in $H\rightarrow WW^{*}$ is not trivial.

\section{Mass Reconsruction in $H\rightarrow WW^{*}\rightarrow\ell^{+}\nu\ell^{-}\bar{\nu}$ }

In the $H\rightarrow WW^{*}\rightarrow\ell^{+}\nu\ell^{-}\bar{\nu}$ channel, there are two neutrinos which escape detection. The system is underconstrained and it is not possible to determine the
momenta of the two neutrinos. Typical analyses in these channels involve
selection criteria on simple kinematic variables, and cross section
upper limits are derived using the distributions of dilepton azimuthal
opening angle which reflects the spin 0 nature of the Higgs boson
\cite{CDFHWW,D0HWW}. To increase the sensitivity of the searches
and to measure the mass, it is desirable to have a variable that has
direct information on the mass of the Higgs boson ($M_{H}$).

There are a couple of kinematic variables that can be used for mass
reconstruction in the $H\rightarrow WW^{*}\rightarrow\ell^{+}\nu\ell^{-}\bar{\nu}$
channel \cite{lesterhwwmass,choihwwmass}. They are either generalizations
of transverse mass ($M_{T}$) or modifications of solutions to kinematic
problens in supersymmetric (SUSY) models. These variables show linear
behavior to $M_{H}$ and are sensitive to it. While these variables
are motivated by kinematics of the event, it is not clear in what sense
these variables are optimal.

We approach the problem of $M_{H}$ measurement from
a different perspective. Since the kinematics of an event reflects
$M_{H}$, we should be able to extract $M_{H}$ from the leptons and
missing transverse momenta which contain the maximal information.
We would like to construct a function such that, on average, $<F(\vec{x}(m))>=m$,
where $\vec{x}$ are the measured quantities from a detector. Since
there are infinitely many such functions, some optimization condition
is necessary to arrive at an optimal function.
This is a problem of non-linear multivariate regression.

There are advanced analyses techniques which allow us to construct
a multivariate function $\hat{F}(\vec{x})$ that can be regarded as
an approximation to the ideal function $F(\vec{x})$ \cite{machinelearning}.
A function is built from a training data set such that $\hat{F}(\vec{x})\approx F(\vec{x})$.
Its performance can be evaluated on a test sample, from variance or some other measure of error.
However, most of these methods are black boxes, such that if we write
down the solution, we would not be able to make much sense of it. Also, these
methods are not able to generalize sufficiently and undesired biases
show up for data sample close to the boundaries of input variables.
Instead, we take an analytic approach to function approximation called
symbolic regression.

\section{Symbolic Regression and Application to $M_H$ Reconsruction}

\subsection{Symbolic Regression}

In a symbolic regression, a function which minimizes certain criteria
(or maximizes fitness) is constructed from the input variables analytically
\cite{naturalevolution,symbregr}. Symbolic regression is 
 an application of genetic programming methodology and genetic algorithms.
Symbolic regression methods are powerful enough to derive invariants,
such as Hamiltonian, from a set of experimental data with non-linearities \cite{science}.
An advantage of symbolic regression is its interpretability, in contrast
to purely numerical methods.

Genetic algorithms are often employed in problems of optimization
with many parameters. It is an application of natural evolution to
computational domain. 
In a genetic algorithm, a set of individuals
form a population, where an individual is represented by a gene. Each
position in a gene can be used to encode some strategy or functionality.
Behavior of an individual, also known as phenotype, is determined by its 
genetic constitution or genotype. The fitness of an individual is evaluated
at the level of phenotype. 

If two individuals with different genotypes show
the same phenotype, then they have identical fitness.  
Fitness is a measure of how well an individual achieves the desired goal.
In genetic algorithms of computational domains, fitness is explicitly defiend, unlike 
the biological world, where fitness is implicit.

Evolution of population is achieved through genetic operations which create new genotypes
or modify existing ones.
Cross-over and mutation operations are genetic operators that can
be used to create individuals for successive generations. Cross-over
operation (sexual reproduction) is applied to a pair of parent genes to create a child gene.
An individual can also undergo a point mutation (asexual reproduction) in a gene. Local minima
in the fitness landscape can be avoided because of this randomness.
Strength of genetic algorithm approach comes
from the randomness of the genetic operations and the variety of genes
present in a population.
Genetic algorithms are finding their way into high energy physics
in optimization problems \cite{garcon,ROOT}. For this study, we created
our own genetic algorithm package to overcome the limitations of existing
tools.

%



\FIGURE{\epsfig{file=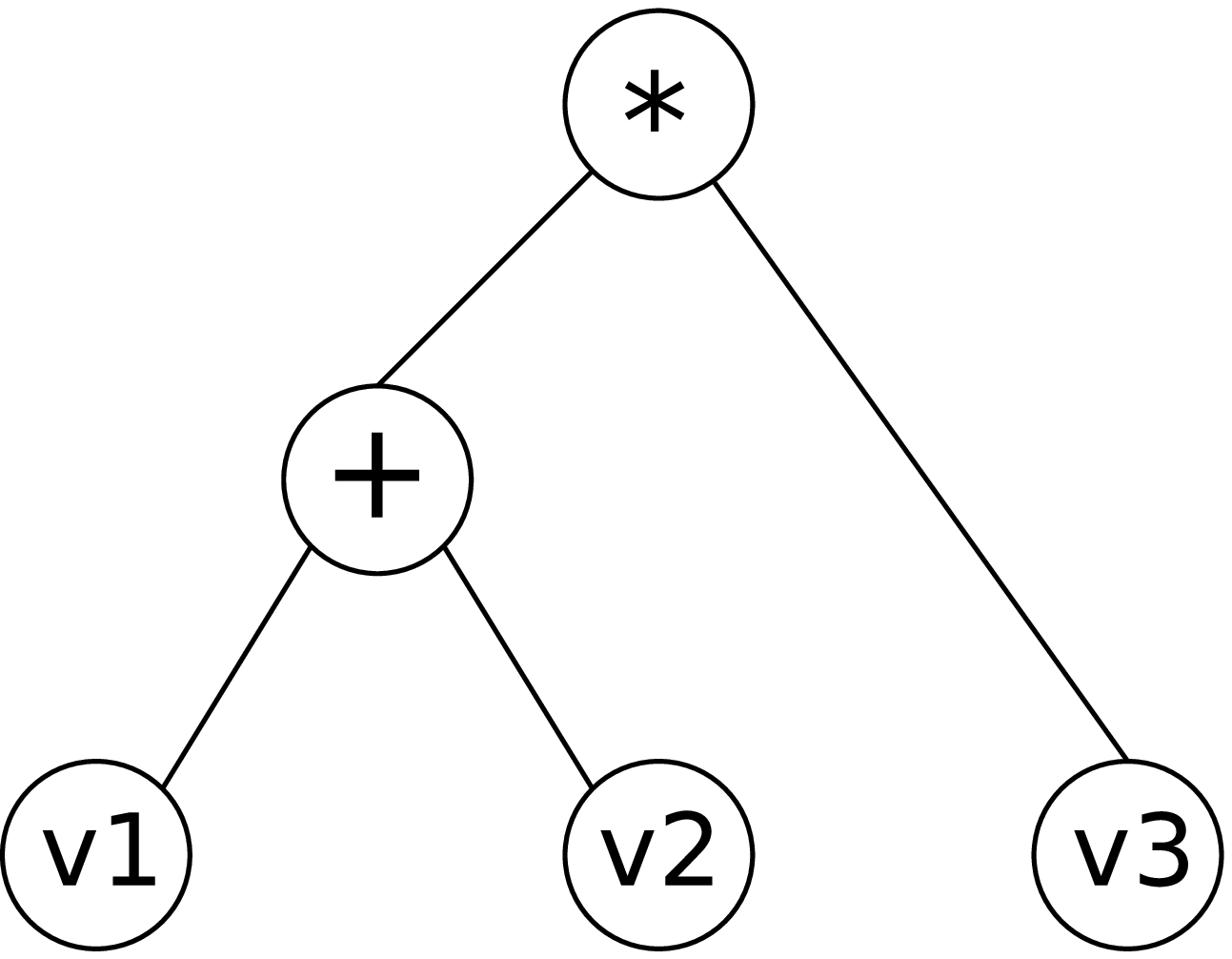,width=0.3\textwidth}
    \caption{A binary tree of expression $(v1+v2)*v3$.}\label{fig:A-binary-tree}}

Symbolic regression is a method that can be applied
to problems where we want to map input variables to the output.
It arrives at the answer using genetic algorithms.
In a symbolic regression, analytic equations form the population. An
individual is usually encoded as an expression binary tree (Fig. \ref{fig:A-binary-tree}),
but other representations are possible. A linear encoding would
make the genes look closer to their biological counterpart, but this
is not necessary. Evaluation of fitness and manipulation of the genes
are much more efficient in a binary tree representation. 
Internal nodes of a binary tree are operators or functions
and terminal nodes are either numerical constants or variables. Set
of operators, variables, constants, and fitness function or minimization
criteria must be defined for each problem.

An initial population is built randomly from a given set of
operators, variables and constants. Individuals of subsequent generations
are created by applying either gene cross-over operations to a pair
of ``father'' and ``mother'' equations to yield a ``child''
equation (Fig. \ref{fig:xoverop}) or through
mutation on existing expressions. Selection
of parents can vary among  implementations. In this study, 
each parent is selected through tournaments.
A tournament is held among a small randomly selected pool from the
population and the best-fit individual is chosen. Through tournaments,
fitter individuals have a greater chance to pass on parts of the genes.

Point mutation operation is applied to each individual nodes randomly
with small probability. This is independent of the sexual reproduction. 
Point mutation mimics random mutations that occur in biological processes. 
The effect of mutation is diversification of the gene pool. Although random
mutations may make the individual less fit, it may still be beneficial when
an offspring inherits some of the mutations. 

At each generation, individuals are sorted according to their fitness and those with poor
fitness are discarded by keeping the population size constant. The
best fit individuals (``the elite'') are passed along to the next
generation without modification, but they can participate in sexual
reproduction. The number of generations or termination criteria has to be 
decided upon as a parameter of the algorithm. The details of 
implementation are discussed in detail elsewhere \cite{inprep}.

\EPSFIGURE{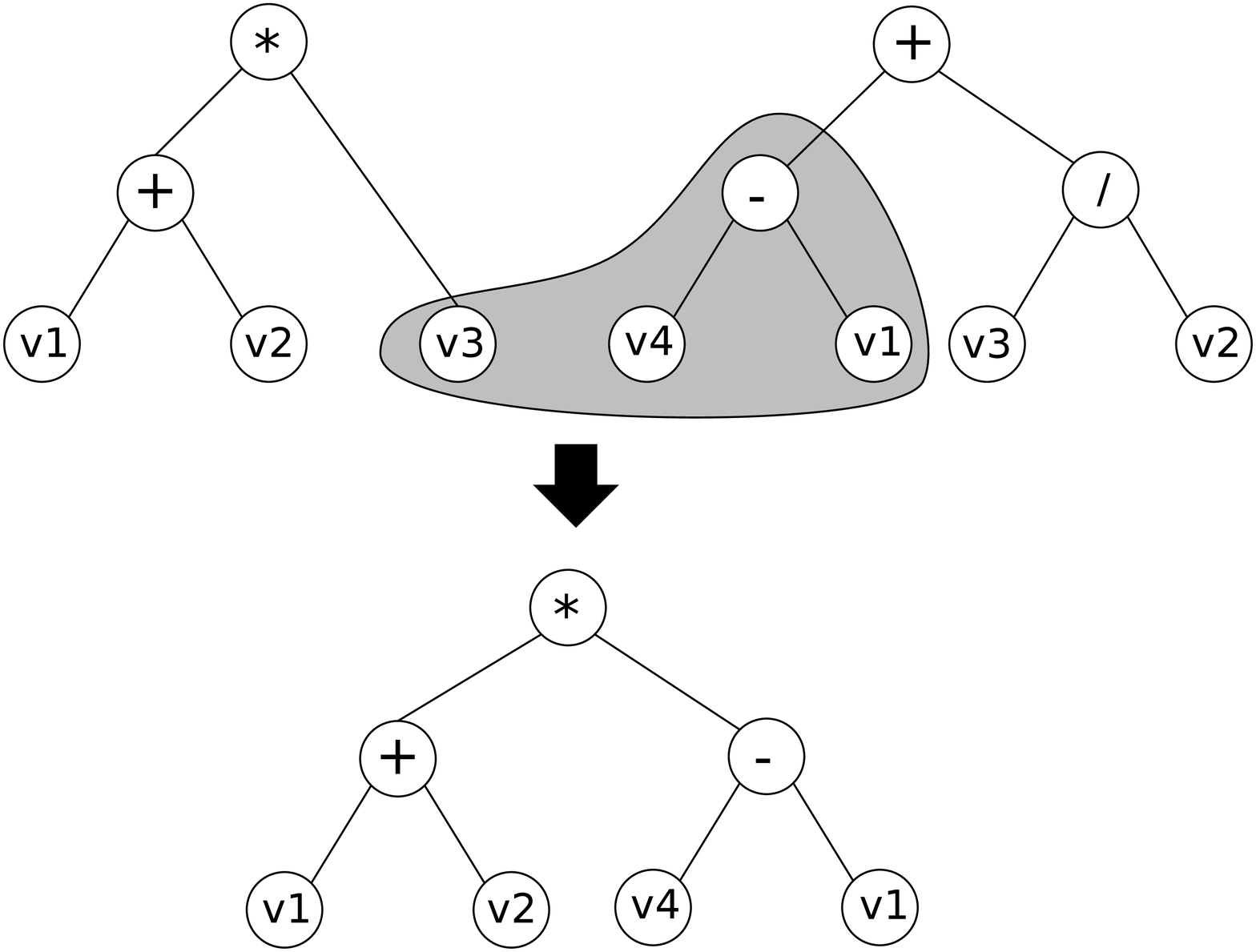, width=0.5\textwidth}{\label{fig:xoverop}
Creation of a child through cross-over operation.
The cross over of genes occuring between $v3$ and $v4-v1$ from the
two parents yields a new child $(v1+v2)*(v4-v1).$ }



In a more traditional method of optimization, a minimum is reached
by descending the fitness landscape in a smooth manner through incremetal
changes. In a genetic algorithm, genetic operations introduce local
changes in the genes, but the behavior of the child 
can be quite different (Fig \ref{fig:xoverop}). It is understood
that fitness landscape can be probed more globally with genetic algorithms.
Maintaining genetic diversity is crucial to the success since
genetic algorithms can still be trapped in local minima if there is
not much genetic diversity. Applying a strong selection pressure on
the population, such as having a large fraction of the population
participating in a tournament, is not necessarily beneficial since
it can effectively reduce the genetic pool to that of a few fit individuals. 

Physical dimensions of resulting formulae of symbolic regression
may not be correct. This is also true of traditional multivariate regression
algorithms. However, in a symbolic regression, we can control the
terms that can appear in an expression. In this study, we created
dimensionally constrained symbolic regression (DCSR) where only terms
that are dimensionally correct can appear. For example, in a DCSR,
terms like $p_{x}+p_{y}^{3}/(p_{x}\cdot p_{z})$ can present, but not
$p_{x}+p_{y}^{3}$. In a DCSR, cross over operations can only occur
among branches with the same physical dimension. 

In tests of simple
problems where we know the optimal answer, such as mass determination
in $W\rightarrow\ell\nu$, DCSR, as well as the normal symbolic regression,
is able to arrive at an equation that differs from the well-known transverse
mass ($M_{T}$) by a multiplicative factor. However, for more complicated
problems, solutions of DCSR converge much more rapidly. And in some cases,
only DCSR produces satisfactory solutions.

\subsection{Symbolic Regression Applied to $H\rightarrow WW^{*}\rightarrow\ell^{+}\nu\ell^{-}\bar{\nu}$ }

Higgs mass determination in $H\rightarrow WW*\rightarrow\ell^{+}\nu\ell^{-}\bar{\nu}$
in hadron colliders is an inportant problem. In this channel,
two lepton momenta $\vec{p}_{\ell1},\vec{p}_{\ell2}$
and the vector sum of the two neutrino transverse momenta $\not\!\!\vec{E}_{T}=\not\!\!\vec{E}_{T\nu1}+\not\!\!\vec{E}_{T\nu2}$
are measured in experiments. 
Since there are only two equations related to neutrino momenta, 
the system is under-constrained. If we knew both $W$
bosons were real, we would still need two extra equations to constrain
the system. Therefore one cannot solve for the neutrino momenta
exactly even in principle. 

Existing studies relied on analysis of kinematics to find expressions
that behave linearly to the Higgs boson mass \cite{lesterhwwmass,choihwwmass}.
In this study, we approach the problem as that of finding an expression
that not only shows linear behavior, but also whose widths of the
mass distribution are narrow. 

Symbolic regression is applied to a data generated with PYTHIA $pp\rightarrow H\rightarrow WW^{*}\rightarrow\ell^{+}\nu\ell^{-}\bar{\nu}$
at $\sqrt{s}=14$ TeV with $M_{H}$ varying from 120 GeV to 200 GeV
\cite{PYTHIA}. Detector simulation is not applied to the data. Momentum
components and energy of the two charged leptons ($p_{1T},p_{1x},p_{1y},p_{1z},E_{1},p_{2T},p_{2x},p_{2y},p_{2z},E_{2}$)
and missing $E_{T}$ information\linebreak ($\not\!\!E_{T},\not\!\! E_{x},\not\!\! E_{y}$)
are used as input variables for the symbolic regression. The fitness
function used is the average of fractional absolute difference: $\frac{1}{N}\sum_{i}|M_{rec,i}-M_{H,i}|/M_{H,i}$. 

Without DCSR, the symbolic regression is not able to yield meaningful results. 
This seems to be due to the larger number of variables used.
The number of terms of dimension 2 with only multiplication
allowed is 78. In our implementation, four basic arithmetic operators ($+,-,\times,\div$)
   and transcendental functions ($\sin,\cos,\log,\exp$) are allowed,
which makes the number of possible terms infinite. 

If fractional root mean-squared (RMS) ($\frac{1}{N}[\sum_{i}(M_{rec,i}-M_{H,i})^2/M_{H,i}^2]^{1/2}$) were used as the fitness function, the symbolic regression 
would get trapped into local minima even with DCSR. 
This is consistent with what is known about genetic
algorithms since outliers pay a heavy penalty with such a fitness function. 
Overall, it has the effect of reducing the diversity. 

Figure \ref{fig:masspred}
shows evolution of fitness of best-fit individuals in 100 runs as
a function of the number of generations. DCSR is able to converge
on meaningful results and yields the best estimate for the $M_{H}^{2}$
as \[
S_{mass}^{2}=2p_{1T}^{2}+2p_{2T}^{2}+3\left(p_{1T}p_{2T}+\not\!\! E_{T}(p_{1T}+p_{2T})-\not\!\!\vec{E}_{T}\cdot(\vec{p}_{1T}+\vec{p}_{2T})-2\vec{p}_{1T}\cdot\vec{p}_{2T}\right).\]
Symmetry of the two leptons in the system is recognized by the symbolic
regression automatically, even though symmetry condition was not imposed.

\FIGURE{
\epsfig{file=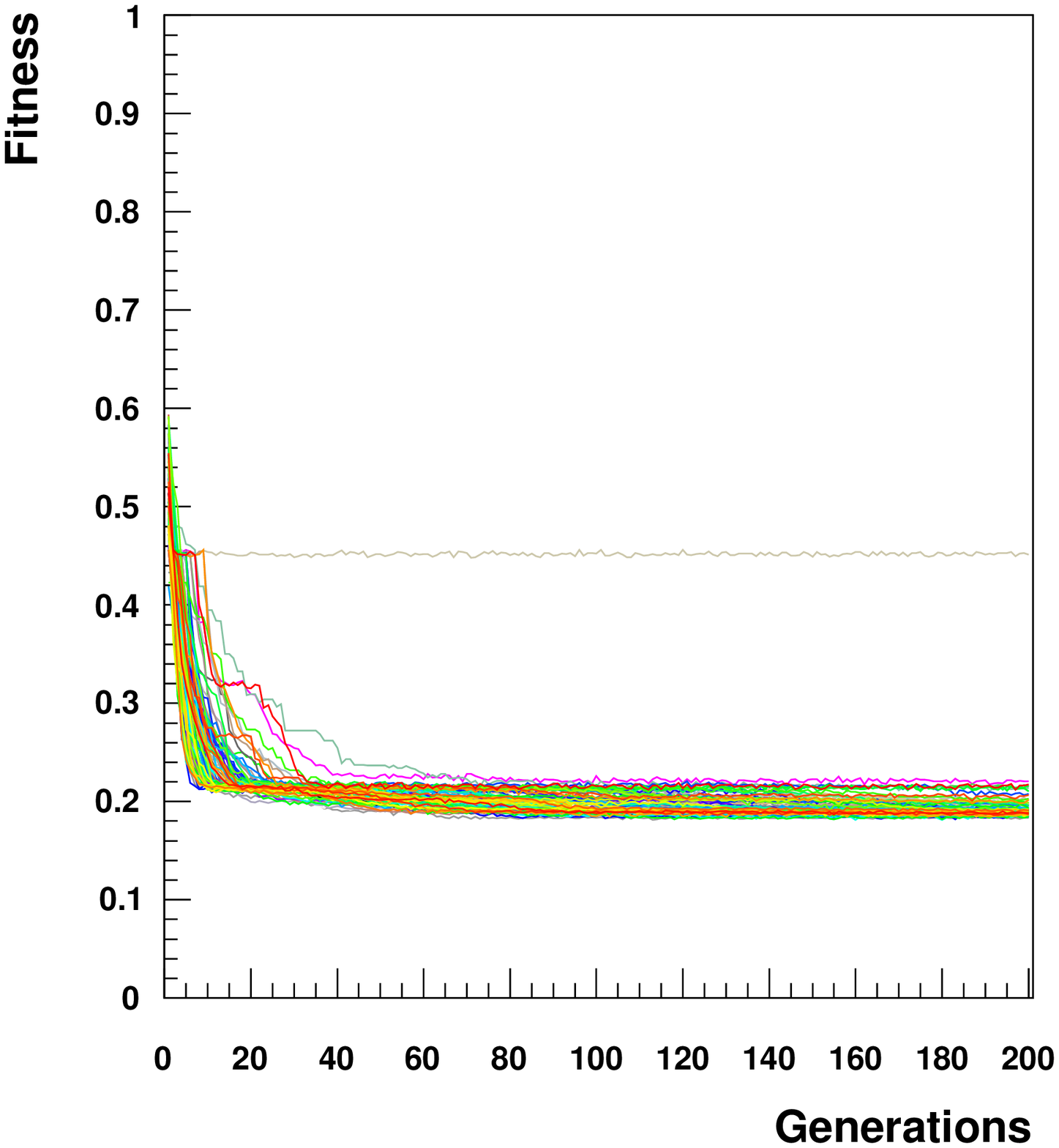,width=0.4\textwidth}
\epsfig{file=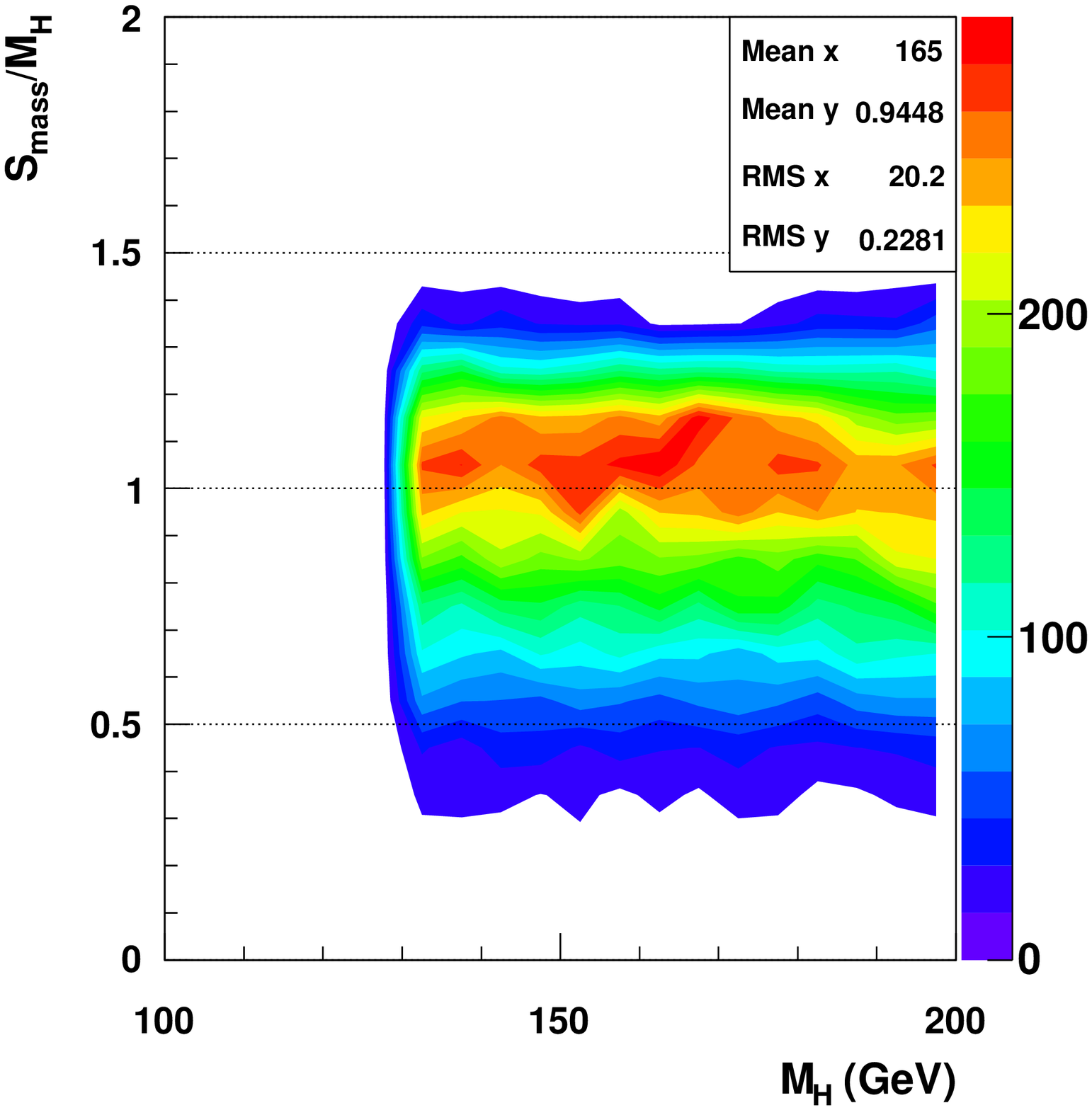,width=0.4\textwidth}
\caption{\label{fig:masspred} Left: Evolution of best fit individuals function
in 100 runs. Right: Distribution of ratio of predicted mass to true
mass, $m_{pred}/m_{H}$, versus the true Higgs mass $m_{H}$. The
sample was generated using PYTHIA.}
}




%
\begin{figure}
\begin{centering}
\includegraphics[width=0.4\textwidth]{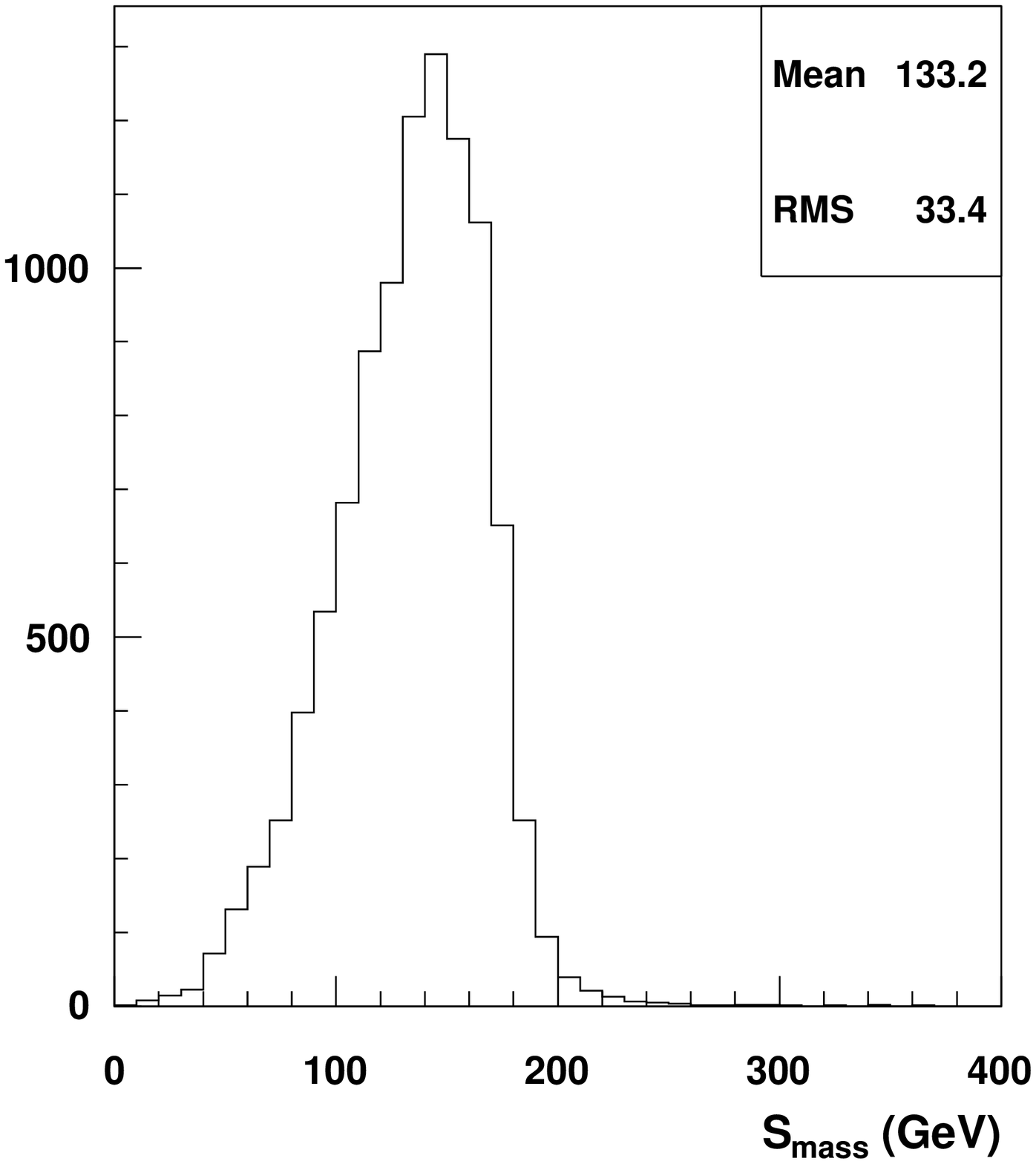}
\par\end{centering}

\caption{\label{fig:mh140}Distribution of reconstructed mass for $S_{mass}$(top)
for $M_{H}=140$ GeV for $pp$ collisions at $\sqrt{s}=14$ TeV.}

\end{figure}

For a Higgs of $M_{H}=160$ GeV which decays to two real $W$ bosons,
if the charged leptons both travel in the same direction, transverse
to the beam with 0 longitudinal momentum components, $S_{mass}=\frac{5}{4}M_{H}$.
The other extreme case, where $\not\!\!\vec{E}_{T}=0$ and the two
lepton momenta are opposite to each other in the transverse plane,
$S_{mass}=\frac{1}{4}M_{H}$. Other configurations of lepton momenta
and $\not\!\!\vec{E}_{T}$ yields different values of $S_{mass}$.
Since we are assuming perfect knowledge on lepton momenta and $\not\!\! E_{T}$,
the width of the distribution reflects the fact that some of the information
on two neutrinos is irretrevably lost. The distribution
of $S_{mass}$ shows good fractional RMS (Fig. \ref{fig:mh140}).
Mass resolution depends not only on the RMS but also on the shape
of the distribution, and this is described in a latter section
under more realistic conditions. By replacing $p_{1T}p_{2T}$ with
$2p_{1T}p_{2T}$ in $S_{mass}$, one can get a variable with the mean closer to
$M_H$, but its distribution is wider.

\begin{figure}
\begin{centering}
\includegraphics[width=0.4\textwidth]{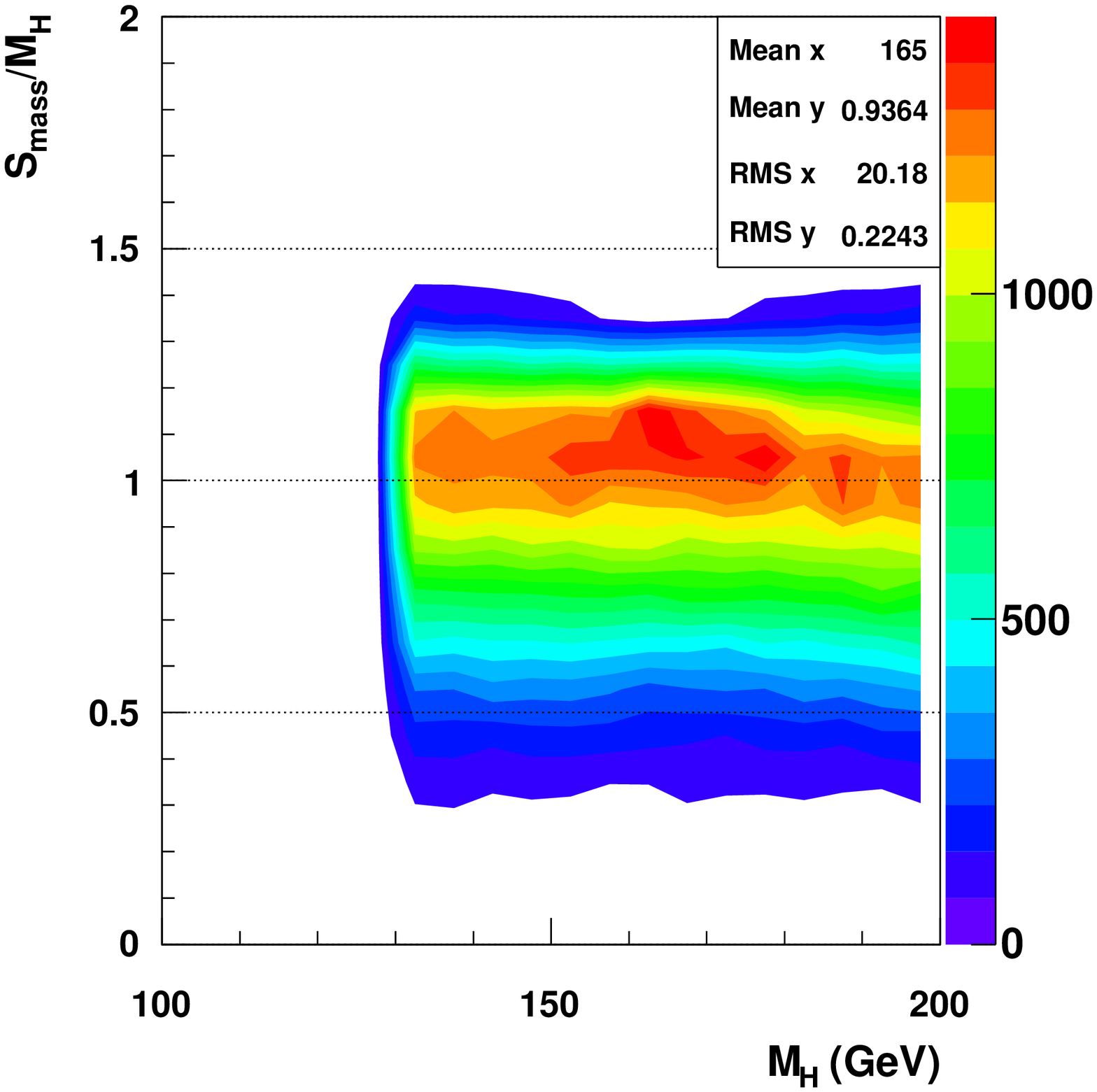}\includegraphics[width=0.4\textwidth]{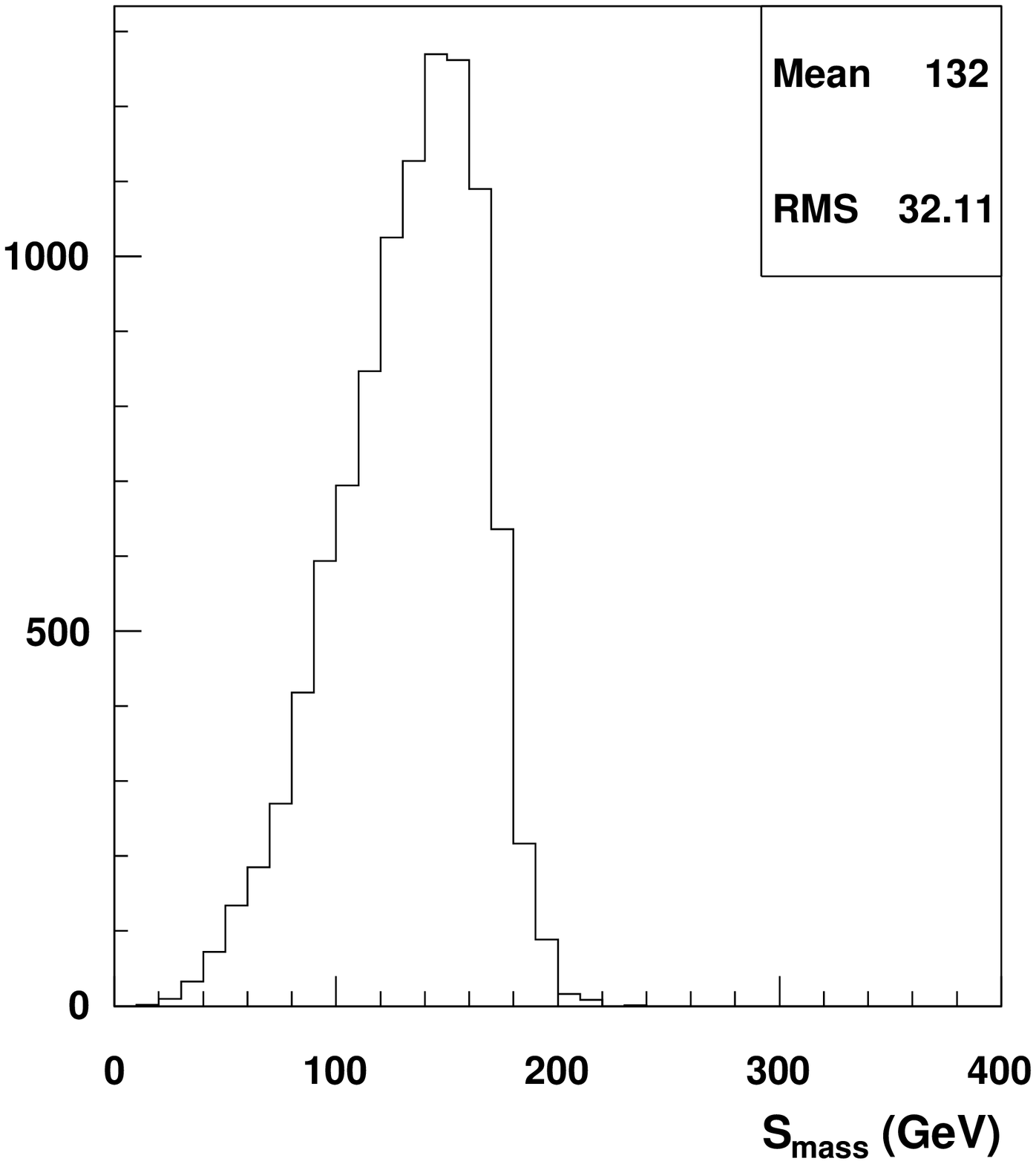}
\par\end{centering}

\caption{\label{fig:hwwtev}Left: the linearity of the $S_{mass}$ 
 in $p\bar{p}\rightarrow H\rightarrow WW^{*}\rightarrow\ell^+\nu\ell^-\bar\nu$
at Tevatron energies $\sqrt{s}=1.96$ TeV. Right: $S_{mass}$ distribution for $M_{H}=140$. }

\end{figure}

Since the simulated data used to derive the equation was from $pp$
collision at $\sqrt{s}=14$ TeV, it is worth to look at how $S_{mass}$
would perform for a different scenario, such as Tevatron where $p\bar{p}$
collides at $\sqrt{s}=1.96$ TeV. The Higgs bosons at the Tevatron
are expected to be produced with a smaller boost and the kinematics
of the final state particles are different. Fortunately, even in this case,
the $S_{mass}$ variables shows linearity and similar fractional RMS
(Fig. \ref{fig:hwwtev}). Therefore, we conclude that $S_{mass}$
captures genuine features of $H\rightarrow WW^{*}\rightarrow\ell^{+}\nu\ell^{-}\bar{\nu}$
system.

\subsection{Mass Sensitivity of $S_{mass}$ variable at LHC}

Sensitivity of a variable to mass depends on the shape of the mass
distributions for the signals. In order to study the sensitivity of
the mass variables under a more realistic conditions, we generated
$pp\rightarrow H\rightarrow WW^{*}\rightarrow\ell^{+}\nu\ell^{-}\bar{\nu}$
events at $\sqrt{s}=14$ TeV using MadEvent generator with PGS v4
detector simulation and reconstruction \cite{Madgraph}. We assume
$pp\rightarrow WW^{*}$ and $pp\rightarrow t\bar{t}$ as backgrounds.
We generated the simulated signal samples in the mass range between
$120\ {\rm GeV}/c^{2}$ and $200\ {\rm GeV}/c^{2}$ at $2.5\ {\rm GeV}/c^{2}$
intervals. Both the signal and background samples are scaled to the
NLO cross sections by applying appropriate K-factors \cite{HiggsNNLO,HWWNLO,topxsec}. 

\begin{figure}
\begin{centering}
\includegraphics[width=0.4\textwidth]{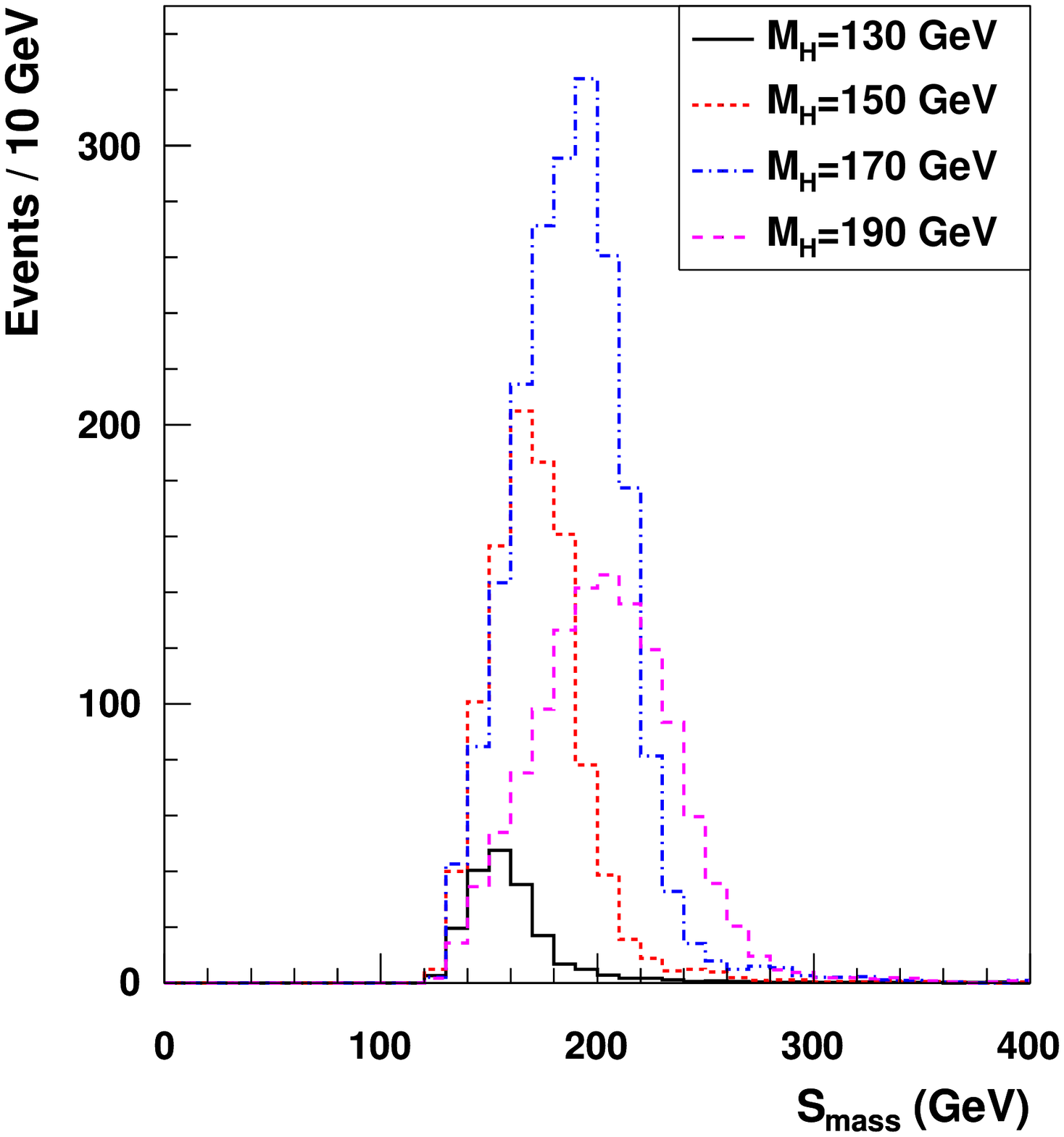}
\par\end{centering}

\caption{\label{fig:smassdistributions}$S_{mass}$ distributions of the Higgs
signal for various $M_{H}$ expected in $10\ fb^{-1}$at LHC after
event selection.}

\end{figure}

The selection critera are as follows:
\begin{itemize}
\item Two leptons of $p_{T}>15$ GeV and $|\eta|<2.5$
\item $12\ {\rm GeV}<M_{\ell\ell}<300$ GeV
\item $\not\!\! E_{T}>30$ GeV
\item $M_{T2}>50$ GeV
\item $|\Delta\phi_{\ell\ell}|<1.8$
\item No hadronic jets with $p_{T}>20$ GeV
\end{itemize}
The last 3 critera reduce the $t\bar{t}$ backgrounds significantly
and $WW^{*}$ backgrounds moderately. The $M_{T2}$ variable
is a good variable to use, since signal-to-background increases \cite{choihwwmass}.
The $S_{mass}$ variable is weakly correlated with the $M_{T2}$ variable, and for larger values
of $M_{T2}$, the $S_{mass}$ distribution becomes sharper. 
The selection has the effect of removing events with smaller values of $S_{mass}$ where backgrounds
are copious.  The $S_{mass}$ distributions
for various values of $M_{H}$ are shown in Fig. \ref{fig:smassdistributions}. 

\begin{figure}
\begin{centering}
\includegraphics[width=0.4\textwidth]{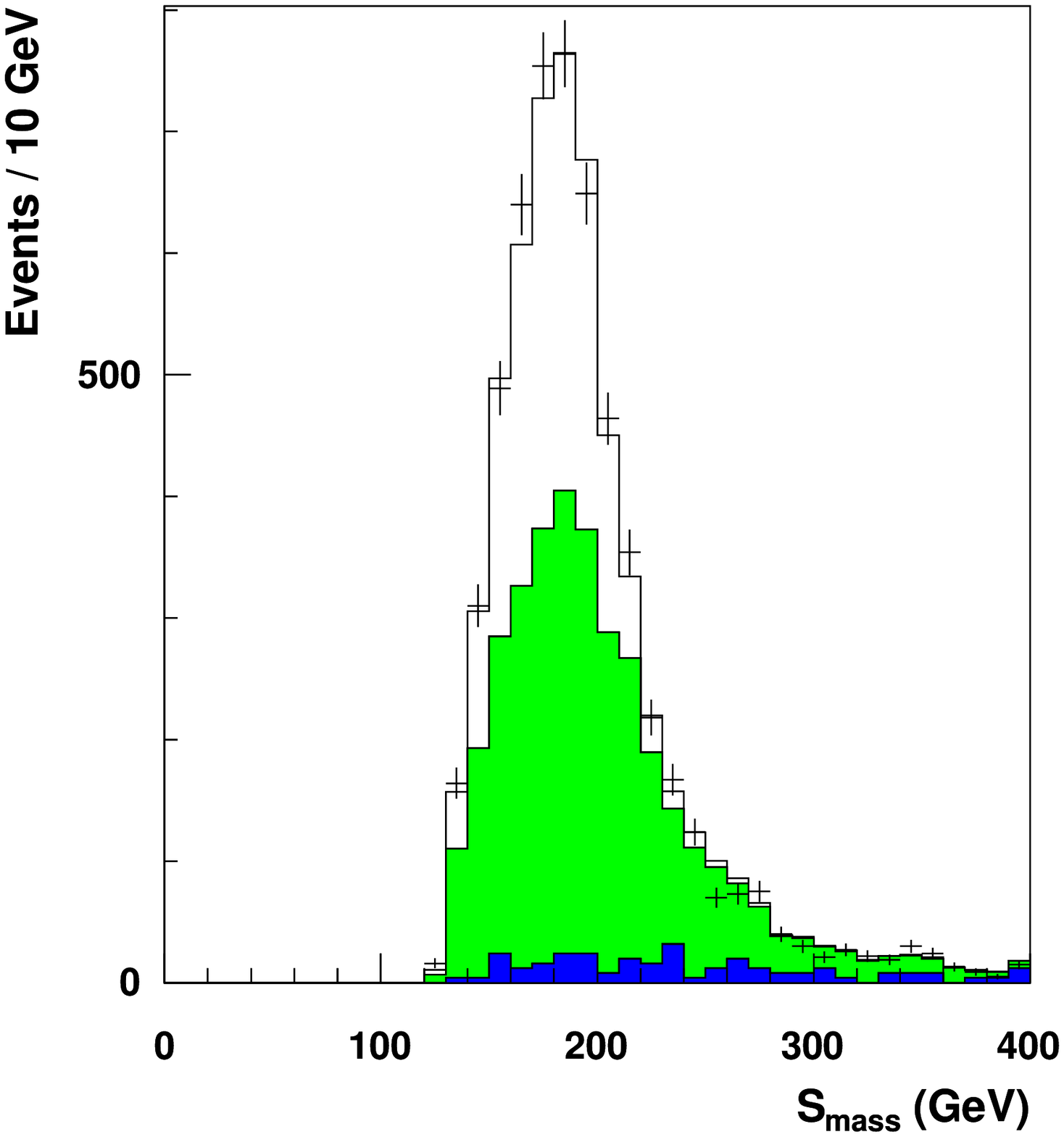}\includegraphics[width=0.4\textwidth]{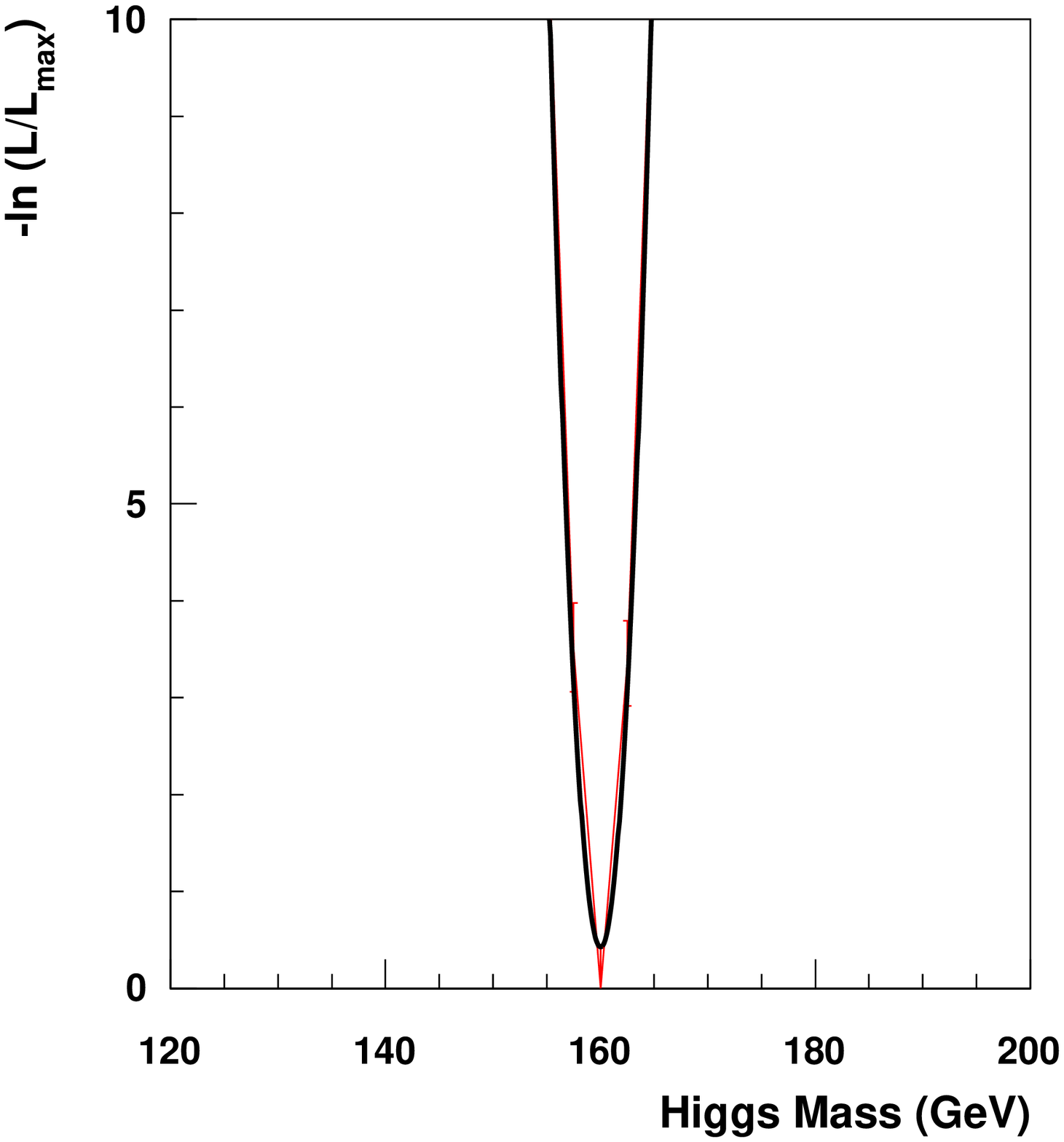}
\par\end{centering}

\caption{\label{fig:Smass-sensitivity}Left: Template histograms of $t\bar{t}$
(dark), $WW^{*}$ (light), and $H\rightarrow WW^{*}$ with $M_{H}=140$
GeV (white) for $10fb^{-1}$. 
The error bars show what a typical data would look
like. Right: Log-likelihood as a function of mass for a 160 GeV Higgs.}

\end{figure}

\begin{figure}
\begin{centering}
\includegraphics[width=0.6\textwidth]{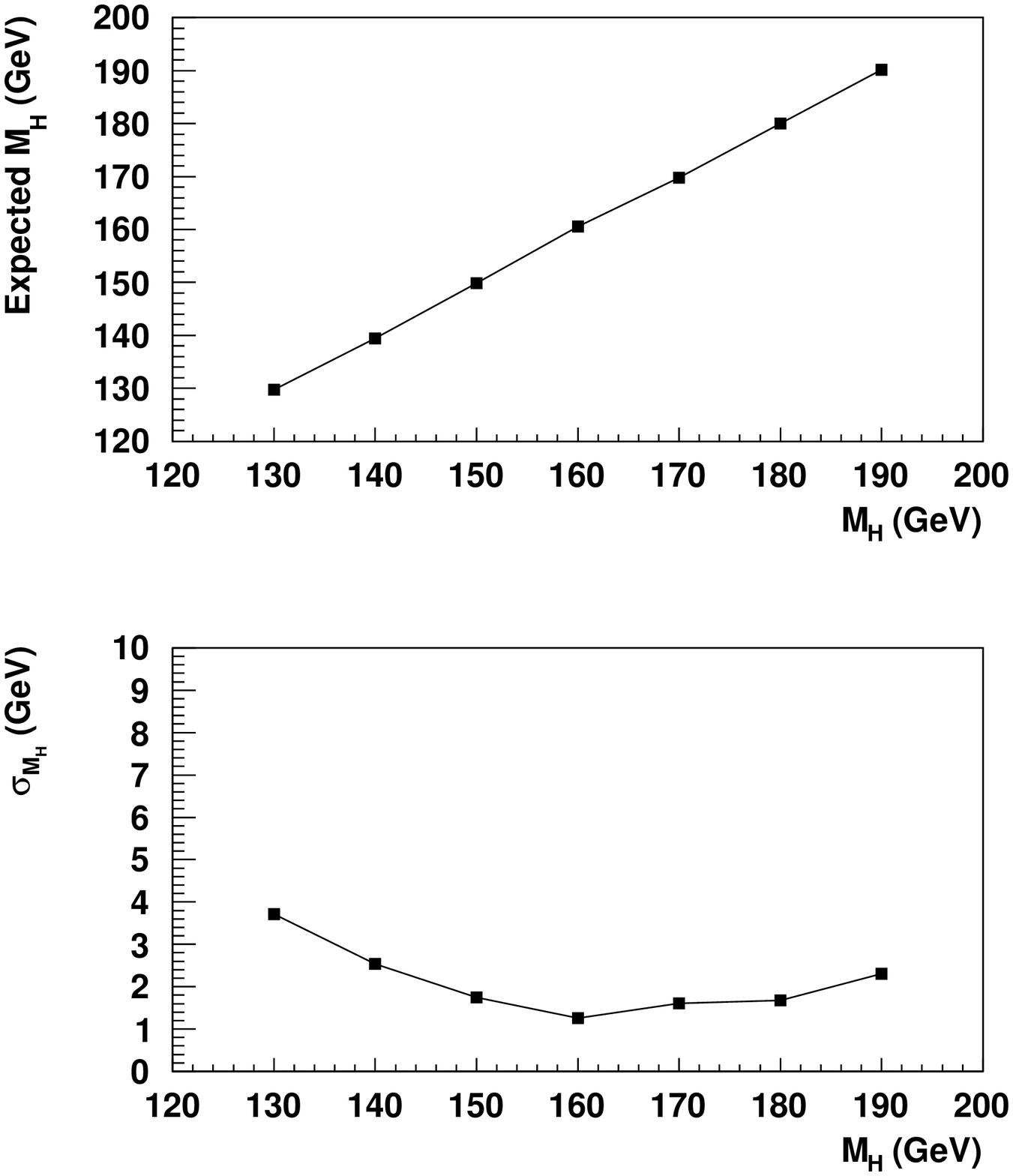}
\par\end{centering}

\caption{\label{fig:Linearity and resolution}Linearity (top) and mass resolution
(bottom) expected as obtained from pseudo-experiments in $10\ fb^{-1}$
for $pp$ collisions at $\sqrt{s}=14$ TeV,}

\end{figure}

To take into account theoretical and experimental uncertainties,
10\% uncertainty in the overall normalization is assumed. 
To evaluate the uncertainties in mass determination, templates of
signal and backgrounds $S_{mass}$ distributions are used to conduct
pseudo-experiments. Log-likelihood is calculated for each mass hypothesis
and then fitted with a parabola to extract the mass resolutions (Fig.
\ref{fig:Smass-sensitivity}). The mass resolution is obtained from
the parabola when $-\ln\mathcal{L}/\mathcal{L}_{max}=\frac{1}{2}$
(Fig. \ref{fig:Linearity and resolution}). Mass resolution improves
from $3.7\ {\rm GeV}$ to $1.3\ {\rm GeV}$ as on-shell decay of Higgs
becomes possible. With $M_{H}$ dependent cuts on $\Delta\phi_{\ell\ell}$
and $M_{T2}$, the mass resoution improves slightly \cite{choihwwmass}.
The $S_{mass}$ variable is correlated with other mass variables for $H\rightarrow WW$, 
but the correlation is not 100\%. Therefore, further improvement may be possible by forming suitable 
combinations of the variables.

\begin{table}
\begin{centering}
\begin{tabular}{c||ccccccc}
\hline 
$M_{H}$ (GeV) & 130 & 140 & 150 & 160 & 170 & 180 & 190\tabularnewline
\hline
$\sigma_{M_{H}}$ (GeV) & 3.7 & 2.5 & 1.8 & 1.3 & 1.6 & 1.7 & 2.3\tabularnewline
\hline
$\sigma_{M_{H}}$ (GeV) optimized & 3.7 & 2.4 & 1.7 & 0.8 & 1.2 & 1.5 & 1.8\tabularnewline
\end{tabular}
\par\end{centering}

\caption{Expected Higgs mass resolutions using $S_{mass}$
variable with $10\ fb^{-1}$ data in $pp$ collisions at $\sqrt{s}=14$
TeV. The last row shows mass resolutions expected for mass-dependent optimized analyses. }

\end{table}

\section{Conclusions}

Symbolic regression is used to derive a kinematic variable which is sensitive
to the mass of the Higgs boson in the $H\rightarrow WW^{*}\rightarrow\ell^{+}\nu\ell^{-}\bar{\nu}$
channel at hadron colliders. With this variable, the mass of the Higgs
boson can be measured with an accuracy of 1 to 4 GeV in the Higgs
mass range between 130 GeV and 190 GeV at the LHC with 10 $fb^{-1}$
of data. This is the first time symbolic regression method has been
applied to high-energy physics problem. 

\begin{acknowledgments}
This work has been supported by Junior Investigator Grant (2009-0069251) of the Korean
National Research Foundation (NRF).
\end{acknowledgments}

\end{document}